\documentclass{elsart}

\usepackage{amssymb}

\begin{document}
\begin{frontmatter}


\begin{flushright}
UWThPh-2004-24\\
UAB-FT-570
\end{flushright}

\title{Violation of a Bell inequality in particle physics experimentally verified?}
\author[ber]{R.A.~Bertlmann},
\author[bra]{A.~Bramon},
\author[gar]{G.~Garbarino},
\author[ber]{B.C.~Hiesmayr}
\address[ber]{Institut f\"ur Theoretische Physik, Boltzmanngasse 5, A-1090 Vienna,
Austria}
\address[bra]{Grup de F{\'\i}sica Te\`orica, Universitat Aut\`onoma
de Barcelona, E-08193 Bellaterra, Spain}
\address[gar]{Dipartimento di Fisica Teorica, Universit\`a di
Torino and INFN, Sezione di Torino, I-10125 Torino, Italy}

\begin{abstract}
Relevant aspects for testing Bell inequalities with entangled
meson--antimeson systems are analyzed. In particular, we argue
that the results of A.~Go, {\em J. Mod. Optics} {\bf 51}, 991
(2004), which nicely illustrate the quantum entanglement of
$B$--meson pairs, cannot be considered as a Bell--test refuting
local realism.
\end{abstract}

\end{frontmatter}

\section{Introduction}

A recent paper \cite{Go} reports that the famous Bell inequality
\cite{Bell} has been tested in a high--energy physics experiment
for the first time. Data on entangled $B$--mesons, produced via
the resonance decay $\Upsilon(4S)\rightarrow B^0\bar B^0$ at the
KEKB asymmetric $e^+ e^-$ collider and collected at the BELLE
detector, were analyzed. Events of EPR--entangled $B^0\bar B^0$
meson pairs, identified via their semileptonic decays, were used
to claim for a violation of a Bell inequality (BI) in the version
of Clauser, Horne, Shimony and Holt (CHSH) \cite{CHSH} by more
than $3\,\sigma$.
In this Letter we analyze the relevant circumstances for testing
Bell inequalities with entangled meson--antimeson systems and,
consequently, the significance of the reported result.

Though the authors widely appreciate the interest in basic
questions of quantum mechanics (QM) explored in particle physics,
they have to argue that the reported result is scarcely relevant
in the discussion of a violation of the Bell inequality for the
entangled $B$--meson system. In the authors opinion, the proof of
the existence of correlations which are stronger than those
explainable by a theory based on the assumptions of {\it locality}
and {\it realism} is not conclusive due to the following two main
drawbacks:
\begin{enumerate}
\item[(1)] ``{\it Active}'' measurements, opening the possibility to choose among
alternative setups, are missing; in other words, there is no free choice for the
experimenter on the the specific question asked to the system.
\item[(2)] The time evolution of an unstable quantum state is unitary only if the state
for the decay products is included. The ``information'' of these
decay products cannot be ignored as done in Ref.~\cite{Go}.
\end{enumerate}

In addition, the authors will briefly review recent proposals
on how to test the peculiar non--local correlations predicted by quantum
theory for EPR--entangled massive systems in high--energy physics.
Most of these proposals refer to two--kaon systems coming from
$\phi(1020)$--resonance decays or proton--antiproton annihilations,
quite similar to the process $\Upsilon(4S)\rightarrow B^0\bar B^0$
considered in Ref.~\cite{Go}.

In 1935 Einstein, Podolsky and Rosen \cite{EPR} claimed to have
shown that quantum mechanics was an incomplete theory. Their
reasoning relied on two assumptions ---realism and locality--- and
on a precise criterion for completeness. The introduction of local
hidden variables, which complement the information contained in
conventional state vectors, would allow for such a completion of
QM. In 1964 John S. Bell \cite{Bell} considered the whole class of
completions with local hidden parameters and showed that all
expectation values or probabilities derived within that class are
constrained to obey certain inequalities. However, expectation
values or probabilities derived within QM can contradict these
Bell inequalities. With this milestone, Bell shifted the original
arguments of Einstein, Podolsky and Rosen about the physical
reality of quantum systems from the realm of philosophy to the
domain of experimental testing. Moreover, a whole new field which
is now of increasing interest was  also opened. Indeed, the
EPR--entanglement is the basic ingredient of new technologies such
as quantum information and quantum communication (see, e.g.,
Ref.~\cite{BertlmannZeilinger}).

In the last two decades we have witnessed an outstanding progress
in testing the peculiar correlations predicted by quantum theory
between outcomes of space--like separated measurements. But a
decisive and loophole--free experiment, which would rule out any
local realistic theory (LRT), has not been yet performed in the
opinion of the authors (for a different view, see
Refs.~\cite{Aspect}). Notwithstanding, the authors want to stress
that they believe ---as it is also the firm consensus in the
community--- that there is almost no doubt that the outcomes of
this type of experiments will agree with QM. The goal of such
discussions and experiments is then to re--educate our intuition
and to understand the very principles of quantum theory such that
one will be able to use them for new technologies.

The deficiencies of experiments testing Bell inequalities are
essentially twofold and usually known as the ``{\it locality}''
and the ``{\it detection efficiency}'' loophole. The Weihs et al.
experiment \cite{Weihs} with entangled photons closes the first
loophole but not the second one. Conversely, the Rowe et al.
experiment \cite{Rowe} with entangled beryllium ions closes the
second loophole but not the one related to {\it locality}. This
{\it locality} loophole requires space--like separated
measurements by the two observers, Alice and Bob, i.e.,
alternative measurement settings  that can be changed sufficiently
fast during the flight of the two particles and that can be chosen
completely at will or (more easily) at random on each side. The
{\it detection efficiency} loophole arises from the low efficiency
of the detectors
---only a small subset of all produced pairs is detected, most of the pairs are lost.
One is then forced to introduce the additional and non--testable fair sampling
hypothesis assuming that the reduced set of detected events behaves like the total
set.

Finally, Hasegawa et.al \cite{Hasegawa} reported an experiment with single neutrons in an
interferometric device which shows a violation of a Bell--like inequality. The
entanglement is achieved not between two separate particles but between two degrees of
freedom of a single neutron, namely, between the path it takes in the interferometer and
its spin component which is different for the two paths. The mathematical description of
the entangled state is the same as for the previously mentioned systems.
However, as there are no two spatially separated particles, it is contextuality rather than
non--locality that is tested (see also Ref.~\cite{BDHH}).

\section{Requirements for testing a Bell inequality and drawbacks of Ref.\cite{Go}}

\subsection{Choice arguments}

Generally, when discussing Bell inequalities and, particularly, in
the CHSH--version, Alice can choose to measure either with setup
$A$ or $A'$, each one having two possible outcomes. Similarly, Bob
can choose between his two dichotomic setups $B$ or $B'$. By her
choice, the particles under Alice's control are projected either
onto the $A$-- or the $A'$--basis, but the measurement outcome in
the chosen basis is, of course, under God's dices and out of
Alice's control. The fact of being able to choose ``{\it
actively}'' between alternative bases is crucial to derive Bell
inequalities from local realism. This possibility of a choice is
strictly needed in order to argue that ``...had Alice chosen in
the very last moment to measure $A$ instead of $A'$, her choice
would not modify the outcome of Bob's measurement...''. As it is
well known \cite{Redhead}, counterfactuality, as exemplified in
the above sentence, is necessary in deriving a genuine Bell
inequality: without an ``{\it active}'' choice it is indeed
impossible to enforce the \emph{locality} condition.

For the previously mentioned experiments with photons, ions or neutrons, such an ``{\it
active}'' choice between alternative measurement bases with dichotomic outcomes was
clearly possible. But the situation is different in the $B$--meson experiment \cite{Go}.
In this case, one starts with the $B^0\bar B^0$ state
\begin{eqnarray}\label{1}
|\psi^{-}(0)\rangle &=&  \frac{1}{\sqrt 2}\big\lbrace\, |B^0\rangle_l\otimes
|\bar{B}^0\rangle_r - |\bar{B}^0\rangle_l\otimes |B^0\rangle_r\big\rbrace\nonumber\\
 &\simeq&
\frac{1}{\sqrt 2}\big\lbrace\, |B_L\rangle_l\otimes |B_H\rangle_r - |B_H\rangle_l \otimes
|B_L\rangle_r\big\rbrace\; ,
\end{eqnarray}
where $l$ and $r$ denote the ``left'' and ``right'' directions of motion of the two
separating $B$--mesons and $|B_{L,H}\rangle = \left\lbrace |B^0 \rangle \pm |\bar
B^0\rangle \right\rbrace /\sqrt 2$, once (small) $\mathcal{CP}$--violation effects are
ignored. One then allows for their time evolution in free space given by
\begin{eqnarray}
\label{2} |B_{L,H}\rangle\, \rightarrow e^{-i m_{L,H} t} e^{-{1
\over 2}\Gamma_{L,H} t} \;|B_{L,H}\rangle\, ,
\end{eqnarray}
where $\Gamma_{L} \simeq \Gamma_{H} = \Gamma_B = 1 /\tau_B \;
(\hbar =1)$ is the common decay width of the  light-- ($m_L$) and
heavy--mass ($m_H$) eigenstates, $B_L$ and $B_H$, of the
non--Hermitian, ``effective--mass" Hamiltonian. The mass
difference, $\Delta m \equiv m_H - m_L$,  induces $B^0$--$\bar
B^0$ oscillations in time, detectable by $B$--meson flavor
measurements with quantum number ``{\it beauty}'' $B = + 1$ for
$B^0$ and $B=-1$ for $\bar B^0$. This requires the discrimination
of the $B^0$ decay modes from their corresponding charge conjugate
modes from $\bar B^0$, e.g., $B^0 \to D^{*-} l^+ \nu$ vs $\bar B^0
\to D^{*+} l^- \bar \nu$, as done in Ref.~\cite{Go}.  As explained
in Refs.~\cite{GoGisin,BN,BH1}, the different times $t_A,
t_{A'},...$ (Alice's side) and $t_B, t_{B'},...$ (Bob's) of the
joint flavor measurements play then the same role as the distinct
orientations of the polarization analyzers in photonic
experiments. The procedure is formally analogous to that in the
above mentioned experiments, but it is by no means an ``{\it
active}'' measurement. There is no way for the experimenter to
force a $B$--meson to decay at a given instant $t_A$ or $t_{A'}$,
i.e., she/he cannot choose ``{\it actively}'' the measurement
bases and the decay just occurs according to the well--known
probabilistic law. It is Nature that decides the measurement bases
leaving no room for counterfactual considerations. Thus, a basic
condition for the correct derivation from LRT of the BI used in
Ref.~\cite{Go} is not fulfilled and the results, despite providing
a notable test of the QM correlations exhibited by $B^0\bar B^0$
entangled pairs, cannot be relevant when confronting LRT vs QM.

\subsection{Unitarity constraints}

But, even if flavor measurements could be {\it actively} induced
at different times, another drawback affects these kind of
Bell--tests. This second drawback, originated by the postulate of
QM according to the evolution of a closed quantum system is
unitary, is a little more involved but not less important. It
requires the discussion of the time evolution of the meson states,
including the possibility of the decay, a case we certainly do not
have to consider for photons or stable spin--$\frac{1}{2}$
particles. In this paragraph we will explain why the normalization
of the expectation value to the surviving meson pairs, Eq.~(9) of
Ref.~\cite{Go}, is not appropriate. Note that without this
normalization no violation of the Bell--CHSH inequality occurs for
reasons we discuss in the following.

Quite generally, we consider now decaying neutral meson systems such as $B^0\bar B^0$ or
$K^0\bar K^0$ pairs. Because of unitarity of the time evolution, the norm of the total
state must be conserved. This means that the decrease of the norm of the meson state must
be compensated by the increase of the decay product state norm. Thus we describe the
complete time evolution of a meson quantum state through a unitary operator $U(t,0)$ as
follows (see Refs.~\cite{BH1,BeatrixDiss,ghirardi92})
\begin{eqnarray}
\label{timeevolution} |M_{1,2}\rangle \longrightarrow |M_{1,2}(t)\rangle =
U(t,0)\,|M_{1,2}\rangle = e^{-i\lambda_{1,2}t}\;|M_{1,2}\rangle + |\Omega_{1,2}(t)\rangle
\;,
\end{eqnarray}
where $|M_{1,2}\rangle$ represents the eigenstate of the
non--Hermitian, ``effective mass'' Hamiltonian and can be written
as superpositions of the flavor states $|M^0\rangle$ and $|\bar
M^0\rangle$. The exponential evolution of the decaying meson state
is given by the eigenvalues
$\lambda_{1,2}=m_{1,2}-\frac{i}{2}\Gamma_{1,2}$, with $m_{1,2}$
the mass and $\Gamma_{1,2}$ the decay width of the meson
$M_{1,2}$. The state $|\Omega_{1,2}(t)\rangle$ represents the
decay products.

Starting then with an entangled $M^0\bar M^0$ pair, the unitary time evolution also
provides a contribution from the decay product states. These introduce a third possible
experimental outcome which complicates the issue because Bell--CHSH inequalities refer to
dichotomic measurements only.

For decaying systems, it is therefore crucial to formulate the experimental dichotomic
question in accordance with unitarity. The appropriate question on the system when it has
evolved up to time $t$ is ``Are you a meson $M^0$ of a certain flavor $f=+1$ or not?''
---question I. It is clearly different to the question ``Are you a meson $M^0$ with
flavor $f=+1$ or an antimeson $\bar M^0$ with $f=-1$?'' ---question II--- as treated in
Ref.~\cite{Go}, since all decay products (an additional information from the quantum
system) are ignored by the latter. Question I admits just two answers, question II is
dichotomic only if conditioned to the survival of both mesons.

\newpage
Let us be more concrete and consider the expectation values for a series of
correlation measurements in these two cases:

\vspace{0.5cm}

\begin{enumerate}
\item[(i)] For question II: ``Are you a meson $M^0$ or an antimeson $\bar M^0$?''
\begin{eqnarray}\label{Enonunitary}
E^{\rm{non-unitary}}(t_l;t_r) = -\cos(\Delta m\Delta t)\cdot
e^{-\Gamma(t_l+t_r)}\;,
\end{eqnarray}
where $\Delta m = m_1-m_2, \Delta t=t_l-t_r$ and
$\Gamma = (\Gamma_1+\Gamma_2)/2\,$. \\

\item[(ii)] For question I: ``Are you a meson $M^0$ or not?''
\begin{eqnarray}\label{Eunitary}
&&E^{\rm{unitary}}(t_l;t_r)=
-\cos(\Delta m \Delta t)\cdot e^{-\Gamma (t_l+t_r)}\nonumber\\
&&\hphantom{E^{\rm{uni}}} +\frac{1}{2}(1-e^{-\Gamma_1 t_l})(1-e^{-\Gamma_2
t_r})+\frac{1}{2}(1-e^{-\Gamma_2 t_l})(1-e^{-\Gamma_1 t_r})\;.
\end{eqnarray}
\end{enumerate}
\vspace{0.5cm}

The second expectation value, Eq.~(\ref{Eunitary}), compared to
the first one where the decay components are ignored, contains
additional terms which express the characteristic contribution
coming from the decay product states $|\Omega_{1,2}(t)\rangle$.

\vspace{0.5cm}

The expectation values of any LRT have to satisfy the following
Bell--CHSH inequality \cite{CHSH}:
\begin{eqnarray}\label{CHSHinequality}
S \;=\;\left|E(t_A;t_B) - E(t_A;t_{B'})\right| + \left|E(t_{A'};t_{B}) +
E(t_{A'};t_{B'})\right|\; \leq\; 2\;.
\end{eqnarray}

However, the calculation of $S$ using quantum mechanical expectation values shows a
critical dependence on the ratio $x = \Delta m/\Gamma$, which can be formulated in the
following way (see also Refs.~\cite{BH1,BeatrixDiss,ghirardi92}).

\vspace{0.9cm}

\textbf{Proposition:} \textit{The unitary expectation values (\ref{Eunitary}) do not
violate the Bell-CHSH inequality for any choice of the four involved times iff \mbox{$x
=\Delta m/\Gamma < N_I$;} the non-unitary ones (\ref{Enonunitary}) do not violate the
inequality iff $x < N_{II}$}.

\vspace{0.5cm}

$N_I$, $N_{II}$ are bounds which we determine numerically. The
values are: $N_I = 2.6$ and $N_{II} = 2.0$ for the $B^0 \bar B^0$,
$D^0 \bar D^0$ and $B^0_s \bar B^0_s$ systems, while for the
$K^0\bar K^0$ system we have $N_I \approx N_{II} = 2.0$ since we
can neglect the width of the long--lived K-meson as compared to
the short--lived one, $\Gamma_L \ll \Gamma_S$, implying
$E^{\rm{non-unitary}}(t_l;t_r) \approx E^{\rm{unitary}}(t_l;t_r)$.

\newpage
The experimental $x$ values for different meson systems are the
following ones:
\begin{center}
\begin{tabular}{|r|c|}
\hline
$x\quad\,$ & $\;$ meson system\\
\hline
0.77 & $B^0\bar B^0$ \\
0.95 & $K^0\bar K^0$ \\
$<\;\;$ 0.03 & $D^0\bar D^0$ \\
$>$ 20.60 & $B^0_s \bar B^0_s$ \\
\hline
\end{tabular}
\end{center}
Therefore, no violation of the Bell--CHSH inequality occurs for the
familiar meson--antimeson systems; only for the last system a violation is
expected.\\

\textbf{R\'esum\'e:} Normalizing the non--unitary expectation
value (\ref{Enonunitary}) to the surviving pairs, $E^{\rm
non-unitary}_{\rm R}(t_l;t_r)=-\cos(\Delta m\, \Delta t)/
\cosh(\Delta \Gamma\, \Delta t/2)$, with $\Delta \Gamma=
\Gamma_1-\Gamma_2$, as in Refs.~\cite{Go}, one obtains a formal
violation of the Bell--CHSH inequality. But this is hardly
relevant for testing LRT versus QM. The reasons are twofold:
Firstly, ``active'' measurements are missing, therefore an
essential hypothesis for the derivation of a genuine Bell
inequality is not satisfied; secondly, the unitary time evolution
of the unstable quantum state
---the decay property of the meson--- is ignored, which is part of its Nature. Therefore
one has to use the unitary formula (\ref{Eunitary}), which,
however, does not lead to a violation of the Bell--CHSH inequality
for the familiar systems: $B^0\bar B^0$, $K^0\bar K^0$, $D^0\bar
D^0$.

\section{Outlook}

It turns out that quantum mechanical tests of meson--antimeson
systems are more subtle than naively expected and one has to
involve other features of the mesons, which are characteristic for
such massive quantum systems, like $\mathcal{CP}$ violation or
regeneration of quantum states. For example, neutral kaons exhibit
$\mathcal{CP}$ violation in $K^0\bar K^0$ mixing. It is remarkable
that $\mathcal{CP}$ violation is connected with the violation of a
BI for different $K^0-\bar K^0$ superpositions (i.e., different
quasi-spin states instead of different times) of neutral kaons
\cite{BH1,Uchiyama,BGH2}.

It is also quite interesting that, using the well known
regeneration mechanism of kaons, novel Bell inequalities can be
established \cite{BN,BramonGarbarino1} and tested with $K^0\bar
K^0$ pairs produced at $\Phi$-factories and $p\bar p$-machines.

Finally, we would like to point out that meson--antimeson systems
allow for other tests of QM. A possible approach to investigate
the nature of entanglement is to experimentally determine the
decoherence of entangled meson pairs \cite{BGH1,BG3,BDH1} and thus
the validity of QM. It turns out that decoherence is strikingly
connected to the entanglement loss of common entanglement measures
\cite{BDH1}, e.g., the entanglement of formation or the
concurrence. Moreover, other subtle features of quantum mechanics
such as quantum erasers \cite{SBGH1}, quantitative duality
\cite{SBGH3} or quantitative complementarity \cite{SBGH4} are
interesting phenomena of meson systems, which have been studied
recently.

\vspace{0.3cm}

\textbf{Acknowledgements:} This work has been supported by
EURIDICE HPRN-CT-2002-00311, INFN and Spanish MCyT,
BFM-2002-02588.

\end{document}